\documentclass[10pt,twoside,a4paper]{article} 
\usepackage{rgd21pap}
\begin{document}
\setcounter{page}{1}
%
%
\title{On the Influence of Gravity on the Thermal Conductivity}


\author{
M. Tij$^1$, V. Garz\'o$^2$, A. Santos$^2$\\
$^1$ D\'epartment de Physique, \\Universit\'e Moulay Isma\"{\i}l, 
Mekn\`es, Morocco\\
$^2$ Departamento de F\'{\i}sica, \\Universidad de Extremadura,
E-06071 Badajoz, Spain }

\maketitle 


\section{Introduction}
\label{p5376sec1}
The simplest heat flow problem consists of a gas enclosed between two 
infinite, parallel plates kept at different temperatures. In the continuum 
limit, the Fourier law establishes a linear relation between the heat flux 
 and the thermal gradient, i.e.,  ${\bf q}=-\kappa \nabla T$, where 
 $\kappa$ is the thermal conductivity coefficient.
The continuum description applies  when $\lambda/L\ll 1$ and 
$\lambda/\ell\ll 1$, where $\lambda$ is the mean free path, $L$ is the 
separation between the plates, and $\ell\sim T|\nabla T|^{-1}$ is the 
characteristic length over which the temperature changes.
Nevertheless, exact results from the Boltzmann equation for Maxwell 
molecules and from the BGK model for general interactions~\cite{SG94,SBG86} 
show that the Fourier law still holds for large gradients ($\lambda \sim 
\ell$), provided that the Knudsen number $\mbox{Kn}=\lambda/L$ is small.

In this paper we are interested in studying the influence on the heat flux
of an external 
field $g$ (e.g., gravity) normal to the plates.
We will also assume that the flow velocity vanishes. This absence of 
convection is possible if the Rayleigh number is smaller than its critical 
value ($\mbox{Ra}<1700$) \cite{T88}.
This precludes the existence of the Rayleigh-B\'enard instability, that has 
been studied for dilute gases by other authors \cite{CS92}.
In addition to $\lambda$, $L$, and $\ell$, the presence of gravity 
introduces a fourth characteristic length, namely
$h\sim k_B T/mg$, which represents the vertical distance over which the 
field produces a significant effect.
In ordinary laboratory conditions, $h$ is several orders of magnitude  
larger than $\lambda$ and $\ell$, so that the constitutive equations are not 
affected by the action of gravity.
However, discrepancies with respect to the Navier-Stokes predictions can be 
expected if $h$ is not extremely large.
According to a recent perturbation solution of the 
Boltzmann equation through second order in $g$~\cite{TGS97}, one can estimate 
that the heat flux 
deviates from the Fourier law as much as $10\%$ if $\lambda\sim\ell\sim 0.01 
h$.
The aim of this paper is to go beyond the second order in $g$ by using the
BGK model of the Boltzmann equation.
Specifically, we will obtain the hydrodynamic profiles as well as the 
pressure tensor and the heat flux through sixth order in gravity.

\section{Description of the problem}
\label{p5376sec2}
        Let us consider a dilute gas described by the BGK kinetic 
        equation \cite{BGK}:
\begin{equation}
\label{p5376b1}
\frac{\partial}{\partial t}f+{\bf v}\cdot\nabla f+\frac{{\bf F}}{m}\cdot
\frac{\partial}{\partial\bf v}f=-\nu(f-f_{L}).
\end{equation}
Here, $f({\bf r},{\bf v},t)$ is the one-particle distribution function,
${\bf F}$ is an external force, $m$ is the mass of a particle, $\nu$ is
a collision frequency, and $f_{L}$ is the local equilibrium distribution
function, that is characterized by the local density $n$, the local velocity 
${\bf u}$, and the local temperature $T$, defined as
\begin{equation}
\label{p5376b3}
\{n,n{\bf u},3nk_BT\} =\int d{\bf v} \{1, {\bf v}, m({\bf v}-{\bf u})^2\}f,
\end{equation}
where $k_B$ is the Boltzmann constant.
The collision frequency is proportional to the density and its dependence on
the temperature models the influence of the interaction potential. For 
instance,
for Maxwell molecules $\nu\propto n$, while $\nu\propto n T^{1/2}$ for hard
spheres.

The problem we want to investigate is that of a  gas 
enclosed in a slab between two plates at different temperatures.
We assume the existence of  a stationary state with spatial variation 
along the normal
direction $z$ only and a {\em constant\/} external field ${\bf F}=
-mg\widehat{\bf z}$ along that direction. 
The constant $g$ can be interpreted as the gravitational acceleration.
In addition, we assume that there is no convection, i.e., ${\bf u}=0$.
In order to ease the notation, it is convenient to introduce dimensionless 
quantities. To that end,
we choose an {\em arbitrary\/} point $z_0$ in the bulk as the origin and 
take the
quantities at that point (denoted by a subscript 0)
as reference values.
Therefore, we define
$T^*\equiv T/T_0$, $p^*\equiv p/p_0$, ${\bf v}^*\equiv {\bf v}/v_0$,
$f^*\equiv (k_BT_0/p_0)v_0^3 f$, $g^*\equiv g/v_0\nu_0$, where $p=nk_BT$
is the hydrostatic pressure and $v_0\equiv (k_BT_0/m)^{1/2}$ is a thermal
velocity.
In these units, $g^*=\lambda_0/h_0$, where we define the mean free path as
$\lambda_0=v_0/\nu_0$. 
In the case of the spatial variable $z$, it is convenient to rescale it in 
a nonlinear way that takes into account the local dependence of the collision
frequency. Consequently, we define
\begin{equation}
\label{p5376b6}
s=v_0^{-1}\int_{z_0}^z dz' \nu(z') .
\end{equation}
Thus,  the stationary BGK equation reads
\begin{equation}
\label{p5376b7}
\left(1+v_z^*\partial_s-g^*\frac{T^*}{p^*}D_v
\right)f^*=f_L^*,
\end{equation}
where  $\partial_s\equiv \partial/\partial s$ and $D_v\equiv 
{\partial}/{\partial v_z^*}$.
Furthermore, for the sake of concreteness, we have restricted ourselves to 
the case of Maxwell molecules (i.e., $\nu\propto p/T$).
In the geometry of the problem, the relevant velocity moments are defined as
\begin{equation}
\label{p5376b9}
M_{\alpha\beta}=\int d{\bf v}^*\, {v^*}^{2\alpha}{v_z^*}^\beta f^*.
\end{equation}
In particular, the corresponding moments at local equilibrium are
\begin{equation}
\label{p5376b9bis}
M_{\alpha\beta}^L=\frac{(2\alpha+\beta+1)!!}{\beta+1}p^*{T^*}^{\alpha-1+\beta/
2}
\end{equation}
for $\beta$ even, being zero otherwise.
Conservation of momentum and energy implies that $\partial_s M_{02}=-g^*$ 
and $\partial_s M_{11}=0$.

In the absence of gravitation ($g=0$), Eq.\ (\ref{p5376b7})
has an {\em exact\/} solution~\cite{SBG86} characterized by a constant 
pressure,
$p^*=1$,
and a ``linear'' temperature profile,
$T^*=1+\epsilon s$, that applies to arbitrary values of the reduced
thermal gradient $\epsilon=\lambda_0/\ell_0$.
The velocity moments
are {\em polynomials\/} in both $s$ and $\epsilon$.
Their explicit
expression for $\beta+2(\alpha-1)\geq 0$ is~\cite{SBG86}
\begin{equation}
\label{p5376b10}
M_{\alpha\beta}=(-1)^\beta
\sum_
{\stackrel{r=0}{(r+\beta)\mbox{even}}}
^{\beta+2(\alpha-1)}
\frac{(2\alpha+\beta+r+1)!!(\alpha-1+\frac{\beta+r}{2})!}
{(\alpha-1+\frac{\beta-r}{2})!(\beta+r+1)}\epsilon^r (1+\epsilon s)^
{\alpha-1+(\beta-r)/2}.
\end{equation}
In particular, $M_{11}=-5\epsilon$, which means that the Fourier law 
holds
even for large thermal gradients.

The motivation of this paper is to analyze the influence of  gravitation
on the  profiles and transport properties of the above steady Fourier
flow.
However, the presence of the operator $D_v$ in Eq.\ (\ref{p5376b7}) 
complicates 
the problem
significantly.
A convenient strategy  is to take the pure steady 
Fourier flow corresponding
to a value of $\epsilon$ equal to the actual thermal gradient at the point 
$s=0$
as a reference state. 
Consequently, we
will carry out a perturbation expansion in powers of $g$:
\begin{equation}
\label{p5376b11}
f^*=f^{(0)}+ f^{(1)} g^*+ f^{(2)} {g^*}^2+\cdots,
\end{equation}
\begin{equation}
\label{p5376b12}
M_{\alpha\beta}=M_{\alpha\beta}^{(0)}+M_{\alpha\beta}^{(1)} g^* + 
M_{\alpha\beta}^{(2)} {g^*}^2+\cdots,
\end{equation}
\begin{equation}
\label{p5376b13}
p^*=p^{(0)}+ p^{(1)} g^*+ p^{(2)} {g^*}^2+\cdots,
\end{equation}
\begin{equation}
\label{p5376b14}
T^*=T^{(0)}+ T^{(1)} g^*+ T^{(2)} {g^*}^2+\cdots,
\end{equation}
where $M_{\alpha\beta}^{(0)}$ is given by Eq.\ (\ref{p5376b10}), $p^{(0)}=1$, 
and
$T^{(0)}=1+\epsilon s$.
By definition, $p^{(k)}(0)=T^{(k)}(0)=\left.\partial_s T^{(k)} 
\right|_{s=0}=0$ for $k\geq 1$.
It must be emphasized that the terms of order ${g^*}^k$ are {\em nonlinear\/}
functions of $\epsilon$ since no restriction  to the order on $\epsilon$ 
exists.
\section{Perturbation expansion}
\label{p5376sec3}
In this section we obtain the hydrodynamic profiles
$p^{(k)}$ and $T^{(k)}$, the momentum flux $M_{02}^{(k)}$, and
the heat flux $M_{11}^{(k)}$ through order $k=6$.
Insertion of Eq.\ (\ref{p5376b11}) into Eq.\ (\ref{p5376b7}) yields
\begin{equation}
\label{p5376c2}
f^{(k)}=\sum_{j=0}^\infty (-v_z^*\partial_s)^j\left[f_L^{(k)}+
D_v\sum_{i=0}^{k-1}
\left(\frac{T^*}{p^*}\right)^{(i)}f^{(k-i-1)}\right].
\end{equation}
This is a {\em formal\/} solution, since $f_L^{(k)}$ is a functional of
$f^{(k)}$ through its dependence on the pressure and temperature.
Taking moments in Eq.\ (\ref{p5376c2}), one has
\begin{eqnarray}
\label{p5376c3}
\Delta 
M_{\alpha\beta}^{(k)}&\equiv&M_{\alpha\beta}^{(k)}-M_{\alpha\beta}^{L{(k)}}
=\sum_{j=1}^\infty (-\partial_s)^j M_{\alpha,\beta+j}^{L{(k)}}
-\sum_{j=0}^\infty(-\partial_s)^j\sum_{i=0}^{k-1}\left(\frac{T^*}{p^*}\right)^
{(i)}
\nonumber\\
&&\times
\left(2\alpha M_{\alpha-1,\beta+j+1}^{(k-i-1)}+(\beta+j)
M_{\alpha,\beta+j-1}^{(k-i-1)}\right).
\end{eqnarray}
The fact that $f$ and $f_L$ have the same hydrodynamic moments leads to the
consistency conditions 
\begin{equation}
\label{p5376c4}
\Delta M_{00}^{(k)}=\Delta M_{01}^{(k)}=\Delta M_{10}^{(k)}=0, 
\end{equation}
for any $k$.
In order to convert Eq.\ (\ref{p5376c3}) into an explicit equation that can be 

solved
recursively, we need to know the spatial dependence of $p^{(k)}$ and 
$T^{(k)}$.
It turns out that $p^{(k)}$ is a polynomial of 
degree $k-2$ in $s$ (except $p^{(1)}$, that is linear in $s$),
while $T^{(k)}$ is a polynomial of degree $k+1$, the coefficients being 
nonlinear functions of the reduced thermal gradient $\epsilon$.
Thus, 
at a given order $k$, insertion of these polynomials into Eq.\ (\ref{p5376c3}) 
and application of the consistency requirements (\ref{p5376c4}) allow one to 
determine the unknown coefficients
and the
problem can be  recursively solved.
Notice that, seen as functions of the actual space variable $z$, $p^{(k)}$ 
and
$T^{(k)}$ are much more complicated than just polynomials. In fact, in the 
absence of gravitational force, the relationship between $s$ and $z$ is 
nonlinear:
$z=\lambda_0(s+\frac{1}{2}\epsilon s^2)$. In general, such a relationship 
can be obtained inverting Eq.\ (\ref{p5376b6}) as
\begin{equation}
\label{p5376c5bis}
z=z_0+\lambda_0\int_0^s ds'\, \frac{T^*(s')}{p^*(s')}.
\end{equation}

The above scheme is straightforward
but tedious to carry out. Since all the manipulations are algebraic, they 
render themselves to the use of symbolic programming languages.
In this paper, we have evaluated the perturbation expansion through sixth 
order in the field. Since the expressions of the hydrodynamic profiles 
become progressively longer,
here we only give the
explicit results through order $k=4$:
\begin{equation}
\label{p5376c7}
p^*=1-sg^*-\frac{276}{5}\epsilon^2 s {g^*}^3 
-\frac{1}{5}s\left[\frac{12}{5}\epsilon
\left(112\,973\epsilon^2+30\right)+588\epsilon^2 s\right]{g^*}^4+{\cal 
O}({g^*}^5),
\end{equation}
\begin{eqnarray}
\label{p5376c8}
T^*&=&1+\epsilon s+\frac{1}{2}\epsilon s^2 g^*-\epsilon s^2\left(\frac{66}{5}
\epsilon-\frac{1}{3}s\right){g^*}^2-\epsilon 
s^2\left[\frac{16}{25}\left(6624\epsilon^2+5\right)\right.
\nonumber\\
&&\left.+\frac{346}{15}\epsilon 
s-\frac{1}{4}s^2\right]{g^*}^3
-\frac{1}{5}\epsilon 
s^2\left[\frac{12}{25}\epsilon\left(50\,765\,962\epsilon^2+31\,445\right)
\right.
\nonumber\\
&&\left.+
\frac{2}{15}\left(399\,621\epsilon^2+200\right)s+
\frac{971}{6}\epsilon s^2-s^3\right]{g^*}^4+{\cal O}({g^*}^5).
\end{eqnarray}

Once the hydrodynamic profiles are known, Eq.\ (\ref{p5376c3}) can be used 
to obtain all the velocity moments at a given order. The most relevant moment
is $M_{11}$, which is related to the heat flux, $q_z=(p_0v_0/2)M_{11}$.
Another important quantity is $M_{02}=P_{zz}/p_0$, which measures the 
anisotropy of the pressure tensor ${\sf P}$.
As said above, in the absence of gravitation the Fourier law applies 
exactly, i.e. $q_z^{(0)}=-\kappa \partial T/\partial z=-(5p_0 
v_0/2)\epsilon$, and the pressure tensor is isotropic, i.e. 
$P_{zz}^{(0)}=p_0$.
In order to characterize the deviations from these Navier-Stokes predictions 
due to gravity, we introduce the following reduced quantities:
\begin{equation}
\label{p5376c9}
\Lambda(\epsilon,g^*)=\frac{q_z}{q_z^{(0)}}=1+\sum_{k=1}^\infty 
\Lambda^{(k)}(\epsilon){g^*}^{k},
\end{equation}
\begin{equation}
\label{p5376c10}
\gamma(\epsilon,g^*)=\frac{\left.P_{zz}\right|_{z=z_0}}{p_0}=1+\sum_{k=1}^
\infty 
\gamma^{(k)}(\epsilon){g^*}^{k}.
\end{equation}
The results show that $\Lambda^{(k)}$ and $\gamma^{(k)}$ are polynomials in 
$\epsilon$ of degree $k$ and a defined parity, 
\begin{equation}
\label{p5376c11}
\Lambda^{(k)}(\epsilon)=\sum_{\ell=0}^k \Lambda^{(k)}_{\ell}\epsilon^\ell,
\quad
\gamma^{(k)}(\epsilon)=\sum_{\ell=1}^k \gamma^{(k)}_{\ell}\epsilon^\ell.
\end{equation}
To second order in $g^*$ we get $\Lambda^{(1)}_1=\frac{58}{5}$, 
$\Lambda^{(2)}_0=\frac{16}{5}$, $\Lambda^{(2)}_2=\frac{47\,968}{25}$, 
$\gamma^{(1)}_1=0$, and $\gamma^{(2)}_2=\frac{84}{5}$.
The coefficients $\Lambda^{(k)}_{\ell}$ and $\gamma^{(k)}_{\ell}$ for 
$3\leq k\leq 6$ are given in Tables \ref{p5376table1} and \ref{p5376table2}, 
respectively.
\begin{table}
\begin{center}
\begin{tabular}{ccccc}
\hline\hline
 &\multicolumn{4}{c}{$k$}\\
 \cline{2-5}
 $\ell$&3&4&5&6\\
 \hline
0&--&$\frac{4984}{25}$&--&$\frac{1488\,752}{25}$\\
1&$\frac{36\,788}{25}$&--&$\frac{84\,010\,272}{125}$&--\\
2&--&$\frac{240\,215\,772}{125}$&--&$\frac{1971\,403\,262\,472}{625}$\\
3&$\frac{171\,421\,552}{125}$&--&$\frac{3152\,117\,447\,716}{625}$&--\\
4&--&$\frac{1482\,769\,502\,032}{625}$&--&$\frac{14\,018\,843\,605\,272\,988}{
625}$\\
5&--&--&$\frac{24\,150\,803\,847\,769\,024}{3125}$&--\\
6&--&--&--&$\frac{651\,597\,385\,814\,170\,947\,712}{15\,625}$\\
\hline\hline
\end{tabular}
\end{center}\caption{Coefficients $\Lambda^{(k)}_\ell$ for $3\leq k\leq 6$}
\label{p5376table1}
\end{table}
\begin{table}
\begin{center}
\begin{tabular}{ccccc}
\hline\hline
 &\multicolumn{4}{c}{$k$}\\
 \cline{2-5}
 $\ell$&3&4&5&6\\
 \hline
1&$\frac{24}{5}$&--&$\frac{30\,048}{25}$&--\\
2&--&$\frac{196\,272}{25}$&--&$\frac{860\,948\,304}{125}$\\
3&$\frac{236\,316}{25}$&--&$\frac{2282\,442\,504}{125}$&--\\
4&--&$\frac{1579\,897\,956}{125}$&--&$\frac{43\,675\,137\,132\,624}{625}$\\
5&--&--&$\frac{20\,495\,361\,514\,332}{625}$&--\\
6&--&--&--&$\frac{449\,964\,261\,460\,472\,196}{3125}$\\
\hline\hline
\end{tabular}
\end{center}\caption{Coefficients $\gamma^{(k)}_\ell$ for $3\leq k\leq 6$}
\label{p5376table2}
\end{table}
\section{Discussion}
\label{p5376sec4}
The numerical coefficients appearing in Eqs.\ (\ref{p5376c7}) and 
(\ref{p5376c8}), as 
well as in Tables \ref{p5376table1} and \ref{p5376table2} clearly indicate 
that the 
expansion in powers of $g^*$, Eq.\ (\ref{p5376b11}), is only asymptotic. This 
does not pose a serious problem, at least from a practical point of view, 
except for values of $g^*$ (say, $g^*> 10^{-2}$) that correspond to 
gravitational fields unrealistically large.
As a consequence, only the first few terms in the expansion are useful for 
small values of $g^*$. In the subsequent analysis,  
terms of order ${g^*}^3$ and higher will not be considered. This also allows 
us to make  a closer comparison with results derived from the Boltzmann 
equation for Maxwell molecules \cite{TGS97}.
Notwithstanding, the knowledge of the remaining terms might be useful to 
attempt to resum the infinite series and get the transport properties for 
arbitrary values of $g^*$.

The solution to the Boltzmann equation \cite{TGS97} through order ${g^*}^2$ 
exhibits the same structure as in Eqs.\ (\ref{p5376c7})--(\ref{p5376c10}), so 
that 
only the numerical coefficients differ. As a matter of fact, the coefficient 
$\frac{66}{5}$ appearing in Eq.\ (\ref{p5376c8}) is replaced by 
$\frac{468}{45}$;
in addition, the Boltzmann solution yields $\Lambda^{(1)}_1=\frac{46}{5}$, 
$\Lambda^{(2)}_0=\frac{12}{5}$, $\Lambda^{(2)}_2=503.7$, 
$\gamma^{(1)}_1=0$ and $\gamma^{(2)}_2=\frac{128}{45}$. The 
differences indicate that the influence of gravity is stronger in the BGK 
description than in the Boltzmann one, especially in the case of the 
pressure anisotropy.
In order to carry out a more detailed comparison, it is convenient to 
construct Pad\'e approximants for the generalized thermal conductivity 
coefficient $\Lambda$. More specifically, we consider the approximants 
\begin{equation}
\label{p5376c12}
\Lambda_{[1,1]}(\epsilon,g^*)=\frac{\Lambda^{(1)}(\epsilon)+\left[
{\Lambda^{(1)}}^2(\epsilon)-\Lambda^{(2)}(\epsilon)\right]g^*}
{\Lambda^{(1)}(\epsilon)-\Lambda^{(2)}(\epsilon) g^*},
\end{equation}
\begin{equation}
\label{p5376c13}
\Lambda_{[0,2]}(\epsilon,g^*)=
\left\{1-\Lambda^{(1)}(\epsilon)g^*+\left[{\Lambda^{(1)}}^2(\epsilon)-
\Lambda^{(2)}(\epsilon) \right]{g^*}^2\right\}^{-1}.
\end{equation}
%
%
\begin{figure}[!ht]
\epsfysize=5cm \epsfbox{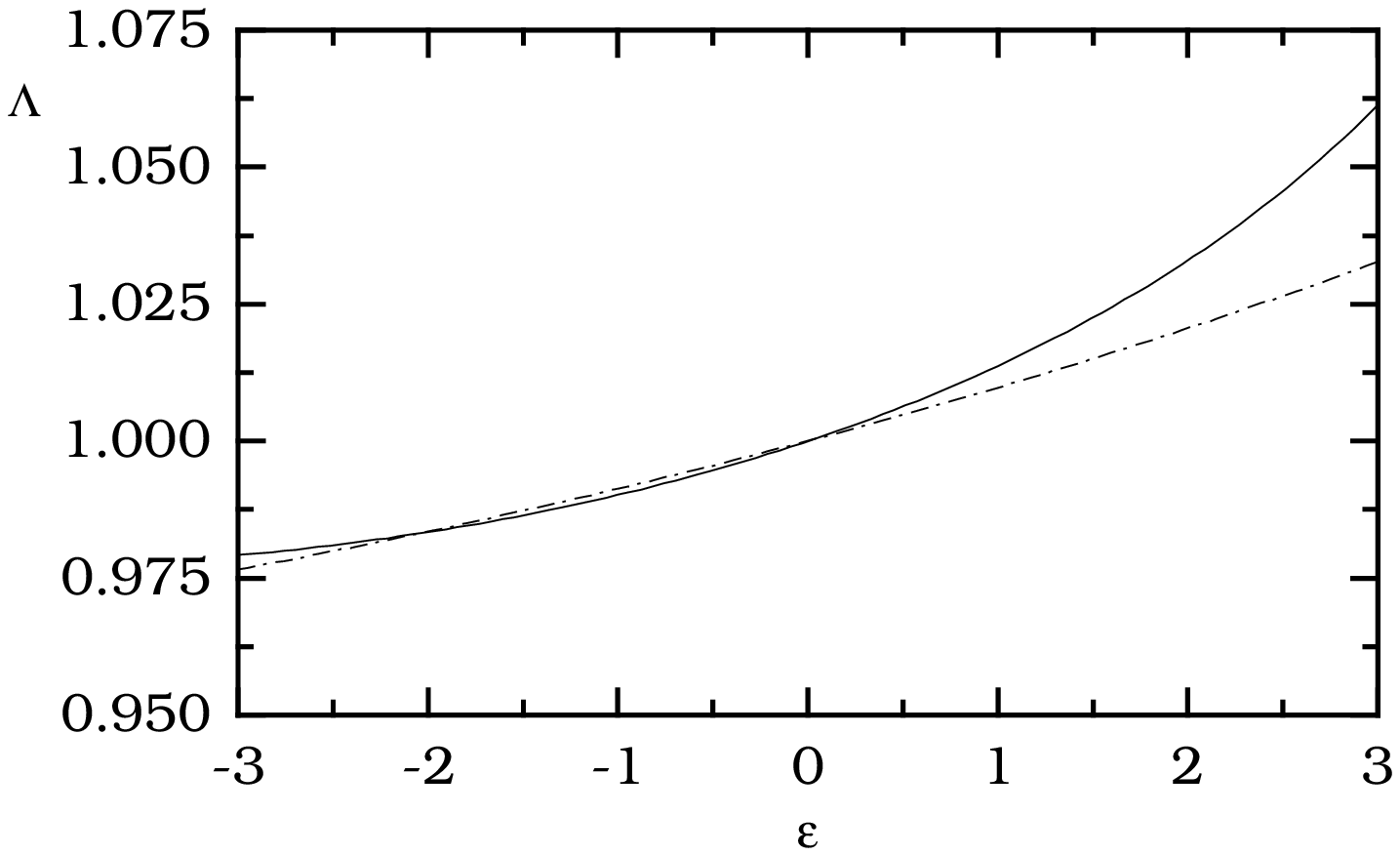}
\caption{Plot of the (reduced) nonlinear thermal conductivity $\Lambda$ 
versus  the thermal gradient $\epsilon$ at $g^*=10^{-3}$, as obtained from 
the BGK model (solid line) and the Boltzmann equation (dashed line).}
\label{p5376fig1} 
\end{figure}
In the case of the BGK equation, these two approximants differ less than 
$2\%$ if $g^*<10^{-2}$ and $|\epsilon| g^*<3\times 10^{-3}$. The differences 
are smaller in the case of the Boltzmann equation. In this range of values 
for $\epsilon$ and $g^*$, a reliable approximation is $\Lambda\simeq 
\frac{1}{2}\left(\Lambda_{[1,1]}+\Lambda_{[0,2]}\right)$.
Figure \ref{p5376fig1} shows this approximation for $\Lambda$ in the interval 
$-3\leq\epsilon\leq3$ at $g^*=10^{-3}$, as given by the BGK and Boltzmann 
equations. 
We observe that the heat flux increases with respect to its Navier-Stokes 
value when one heats from above ($\epsilon>0$), while the opposite happens 
when one heats from below ($\epsilon<0$). This effect is not symmetric, 
since it is more significant if $\epsilon>0$ than if $\epsilon<0$. As said 
above, Fig.\ \ref{p5376fig1} also shows that the influence of gravity is less 
important in the Boltzmann description.

In summary, we have solved the BGK model for a gas simultaneously subjected 
to a thermal gradient and a parallel gravity field of magnitude $g$, in the 
absence of convection. The solution has been obtained by a perturbation 
expansion in powers of gravity, the reference state being the 
non-equilibrium pure Fourier flow with arbitrarily large thermal gradients.
We have explicitly obtained the solution through sixth order in the field.
The results clearly indicate that the expansion is not convergent, although 
it seems to be at least  asymptotic.
The work reported in this paper extends previous results derived from the 
Boltzmann equation to second order in the field 
\cite{TGS97}. The similarity in the structure of the coefficients appearing 
in both descriptions suggests that the expansion obtained from the Boltzmann 
equation is also asymptotic.
Nevertheless, given that at  practical level, the values of $g$ are small, 
the usefulness of the expansion is restricted to the first few terms.

The main results concerning the transport of momentum and energy are that 
the external field induces (i) anisotropy in the pressure tensor, 
$(P_{zz}-p)/p\simeq \frac{84}{5}\epsilon^2  {g^*}^2$, and (ii)
deviations from the Fourier law, $q_z/{q_z^{(0)}}-1 
\simeq \frac{58}{5}\epsilon g^*$.
While the first effect is of second order, the correction to the heat flux 
is of first order, so that it depends on the sign of the thermal gradient.
As a consequence, the heat transport is inhibited when the gas is heated 
from below ($\epsilon<0$), while the opposite happens when the gas is heated 
from above ($\epsilon>0$).
\smallskip

This work has been done under the auspices of the Agencia Espa\~nola de 
Cooperaci\'on Internacional (Programa de Cooperaci\'on Interuniversitaria 
Hispano-Marroqu\'{\i}).
V.G. and A.S. acknowledge partial support from the DGES (Spain) through 
Grant No.\ PB97-1501 and from the Junta de Extremadura (Fondo Social Europeo)
through Grant No.\ PRI97C1041.
The authors are grateful to Prof.\ Y. Sone for valuable discussions about 
the subject of this paper.

\end{document}